\documentclass[%
	final,
	conference,
    comsoc,
	letterpaper,
	oneside,
	twocolumn,
	nofonttune,%
]{IEEEtran}%
\PassOptionsToPackage{usenames,dvipsnames,svgnames,x11names,table,prologue}{xcolor}%
\PassOptionsToPackage{hyphens}{url}%
\PassOptionsToPackage{nomessages}{fp}%
\ifCLASSINFOpdf
  \usepackage[pdftex]{graphicx}
\else
  \usepackage[dvips]{graphicx}
\fi
\ifCLASSOPTIONcompsoc
   \usepackage[caption=false,font=normalsize,labelfont=sf,textfont=sf]{subfig}
\else
   \usepackage[caption=false,font=footnotesize]{subfig}
\fi

\usepackage{stfloats}
\usepackage[english]{babel}%
\usepackage[utf8]{inputenc}%
\usepackage[babel,style=english]{csquotes}%
\usepackage{hyphsubst}%
\usepackage[%
	activate={true,%
	nocompatibility},%
	final,%
	tracking=true,%
	kerning=true,%
	spacing=true,%
	factor=1100,%
	stretch=10,%
	shrink=10%
]{microtype}%
\usepackage{setspace}%
\usepackage{xcolor}%
\definecolor{todonotecol}{RGB}{250,0,0}%
\usepackage{xparse}
\usepackage[%
	colorlinks=false,%
	urlcolor=black,%
	linkcolor=black,%
	citecolor=black,%
	filecolor=black,%
	breaklinks,%
	]{hyperref}%
\usepackage{url}%
\usepackage[%
	acronym,%
	nopostdot,%
	seeautonumberlist,%
	shortcuts,%
	section=chapter,%
	toc,%
]{glossaries}%
\glsaddkey*{url}
{}
{\glsentryurl}
{\Glsentryurl}
{\glsurl}
{\Glsurl}
{\GLSurl}
\glsaddkey*{longpltwo}
{}
{\glsentrylongpltwo}
{\Glsentrylongpltwo}
{\glslongpltwo}
{\Glslongpltwo}
{\GLSlongpltwo}
\newglossaryentry{}{%
	name={},%
	plural={},%
	description={},%
}%
\newacronym{it}{IT}{information technology}
\newacronym{ot}{OT}{operational technology}
\newacronym{opcua}{OPC UA}{Open Platform Communications Unified Architecture}
\newacronym{tsn}{TSN}{IEEE time-sensitive networking}
\newacronym{scada}{SCADA}{supervisory control and data acquisition}
\newacronym{vpc}{VPC}{Virtual Process Controller}
\newacronym{plc}{PLC}{Programmable Logic Controller}
\newacronym{udp}{UDP}{User Datagram Protocol}
\newacronym{pubsub}{PubSub}{Publish Subscribe}
\newacronym{vm}{VM}{virtual machine}
\newacronym{pc}{PC}{personal computer}
\newacronym{ptp}{PTP}{Precision Time Protocol}
\newacronym{gptp}{gPTP}{generalized Precision Time Protocol}
\newacronym{ntp}{NTP}{Network Time Protocol}
\newacronym{splc}{Soft-PLC}{Software-PLC}
\newacronym{rte}{RTE}{Real-Time Ethernet}
\newacronym{osi}{OSI}{Open Systems Interconnection}
\newacronym{mac}{MAC}{Media Access Control}
\newacronym{e2e}{E2E}{end-to-end}
\newacronym{tcp}{TCP}{Transmission Control Protocol}
\newacronym{hmi}{HMI}{Human-Machine Interface}
\newacronym{kpi}{KPI}{Key Performance Indicator}
\newacronym{mes}{MES}{Manufacturing Execution System}
\newacronym{rami4.0}{RAMI 4.0}{Reference Architectural Model Industrie 4.0}
\newacronym{tacnet4.0}{TACNET 4.0}{Tactile Internet 4.0}
\newacronym{3gpp}{3GPP}{3rd Generation Partnership Project}
\newacronym{5gppp}{5GPPP}{5G Infrastructure Public Private Partnership}
\newacronym{itu}{ITU}{International Telecommunication Union}
\newacronym{ngnm}{NGNM}{Next Generation Mobile Networks}
\newacronym{agv}{AGV}{automated guided vehicle}
\newacronym{v2v}{V2V}{vehicle-to-vehicle}
\newacronym{v2x}{V2X}{vehicle-to-infrastructure}
\newacronym{d2x}{D2X}{device-to-x}
\newacronym{qos}{QoS}{quality of service}
\newacronym{noa}{NOA}{NAMUR Open Architecture}
\newacronym{dcs}{DCS}{distributed control system}
\newacronym{m2m}{M2M}{machine-to-machine}
\newacronym{sdn}{SDN}{software-defined networking}
\newacronym{erp}{ERP}{enterprise resource planning}
\newacronym{ip}{IP}{Internet Protocol}
\newacronym{lte}{LTE}{Long Term Evolution}
\newacronym{5g}{5G}{5th generation wireless communication systems}
\newacronym{harq}{HARQ}{Hybrid Automatic Repeat Request}
\newacronym{rat}{RAT}{radio access technology}
\newacronym{mc}{MC}{Multi-Connectivity}
\newacronym{fbmc}{FBMC}{Filter Bank Multicarrier}
\newacronym{gfdm}{GFDM}{Generalized Frequency Division Multiplexing}
\newacronym{sla}{SLA}{service-level agreement}
\newacronym{5qi}{5QI}{5G QoS Indicator}
\newacronym{bmbf}{BMBF}{German Federal Ministry of Education and Research}
\newacronym{nrt}{NRT}{non real-time}
\newacronym{irt}{IRT}{isochronous real-time}
\newacronym{rt}{RT}{real-time}
\newacronym{ar}{AR}{augmented reality}
\newacronym{wlan}{WLAN}{wireless local area network}
\newacronym{cpu}{CPU}{central processing unit}
\newacronym{ram}{RAM}{random-access memory}
\newacronym{ue}{UE}{user equipment}
\newacronym{ran}{RAN}{radio access network}
\newacronym{bs}{BS}{base station}
\newacronym{cn}{CN}{core network}
\newacronym{epc}{EPC}{evolved packet core}
\newacronym{mec}{MEC}{mobile edge cloud}
\newacronym{rtt}{RTT}{round-trip time}
\newacronym{vpn}{VPN}{virtual private network}
\newacronym{vpf}{VPF}{Virtual Process Control Function}
\newacronym{vdi}{VDI}{Association of German Engineers}
\newacronym{ipc}{IPC}{industrial pc}
\newacronym{rtos}{RTOS}{real-time operating system}
\newacronym{mano}{VPC M\&O}{VPC Manager \& Orchestrator}
\newacronym{i/o}{I/O}{Input / Output}
\newacronym{fbd}{FBD}{Function Block Diagram}
\newacronym{ldd}{LDD}{Ladder Logic Diagram}
\newacronym{sfc}{SFC}{Sequential Function Chart}
\newacronym{il}{IL}{Instruction List}
\newacronym{st}{ST}{Structured Text}
\newacronym{os}{OS}{Operating System}
\newacronym{mqtt}{MQTT}{message queuing telemetry transport }
\newacronym{amqp}{AMQP}{advanced message queuing protocol}
\newacronym{api}{API}{application programming interface}
\newacronym{embb}{eMBB}{enhanced mobile broadband}
\newacronym{iot}{IoT}{Internet of Things}
\newacronym{urllc}{URLLC}{ultra-reliable low-latency communication}
\newacronym{mmtc}{mMTC}{massive machine-type communications}
\newacronym{iiot}{IIoT}{industrial IoT}
\newacronym{iiot2}{IIoT}{industrial Internet of Things}
\newacronym{npn}{NPN}{non-public network}
\newacronym{prb}{PRB}{Physical Ressource Block}
\newacronym{gnb}{gNB}{5G base station}
\newacronym{5gs}{5GS}{5G system}
\newacronym{5gc}{5GC}{5G core network}
\newacronym{oai}{OAI}{OpenAirInterface}
\newacronym{enb}{eNB}{4G base station}
\newacronym{sdr}{SDR}{Software Defined Radio}
\newacronym{usrp}{USRP}{Universal Software Radio Peripheral}
\newacronym{mme}{MME}{Mobility Management Entity}
\newacronym{hss}{HSS}{Home Subscriber Server}
\newacronym{pdn}{PDN}{Packet Data Network}
\newacronym{sgw}{SGW}{Serving Gateway}
\newacronym{gtp}{GTP}{GPRS Tunneling Protocol}
\newacronym{mimo}{MIMO}{Multiple Input Multiple Output}
\newacronym{fpga}{FPGA}{Field Programmable Gate Array}
\newacronym{pss}{PSS}{primary synchronization signal}
\newacronym{sss}{SSS}{secondary synchronization signal}
\newacronym{sfn}{SFN}{system frame number}
\newacronym{mib}{MIB}{master information block}
\newacronym{pbch}{PBCH}{Physical Broadcast Channel}
\newacronym{ism}{ISM}{Industrial, Scientific and Medical}
\newacronym{e-utra}{E-UTRA}{evolved UMTS Terrestrial Radio Access}
\newacronym{bldc}{BLDC}{brushless DC}
\newacronym{dl}{DL}{Downlink}
\newacronym{ds-tt}{DS-TT}{Device-Side TSN Translator}
\newacronym{nw-tt}{NW-TT}{Network-Side TSN Translator}
\newacronym{upf}{UPF}{User Plane Function}
\newacronym{tsi}{TSi}{ingress timestamping}
\newacronym{tse}{TSe}{egress timestamping}
\newacronym{gm}{GM}{grandmaster}
\newacronym{sn}{SN}{sequence number}
\newacronym{rbis}{RBIS}{Reference Broadcast Infrastructure Synchronization}
\newacronym{cococo}{CoCoCo}{communication-control codesign}
\newacronym{3c}{3C}{joint design of communications, control, and computing}
\newacronym{cots}{COTS}{commercial off-the-shelf}
\newacronym{ie}{IE}{industrial Ethernet}
\newacronym{ap}{AP}{access point}
\newacronym{ofdm}{OFDM}{Orthogonal Frequency-Division Multiplexing}
\newacronym{csma}{CSMA-CA}{carrier sense multiple access with collision avoidance}
\newacronym{dcf}{DCF}{distributed coordination function}
\newacronym{pcf}{PCF}{point coordination function}
\newacronym{hcf}{HCF}{hybrid coordination function}
\newacronym{edca}{EDCA}{enhanced distributed channel access}
\newacronym{hcca}{HCCA}{\gls{hcf} controlled channel access}
\newacronym{tdma}{TDMA}{time division multiple access}
\newacronym{ofdma}{OFDMA}{Orthogonal Frequency-Division Multiple Access}
\newacronym{mumimo}{MU-MIMO}{multi-user \gls{mimo}}
\newacronym{bi}{BI}{beacon interval}
\newacronym{ssid}{SSID}{service set identifier}
\glsdisablehyper
\usepackage[%
	backend=biber,%
	style=ieee,%
	isbn=false,%
	hyperref=true,%
	maxbibnames=99,%
	sorting=none,%
	natbib=true,%
	language=english,%
	defernumbers=true,%
	]{biblatex}%
\DeclareFieldFormat{sentencecase}{\csname bbx@colon@search\endcsname#1}

\usepackage{nameref}
\usepackage{soul}
\usepackage{flushend}
\usepackage{textcomp}
\usepackage{calc}
\usepackage{xkeyval}
\usepackage{multirow}
\usepackage{tabulary}
\usepackage{makecell}
\usepackage{pgfplots}
\usepackage{tikz}
\usepackage{textcomp} 
\usepackage{verbatim}
\newcommand{\nl}{\par\noindent} 
\newcommand{\mytilde}{{\raise.17ex\hbox{$\scriptstyle\mathtt{\sim}$}}}

\newlength\textheighttemp%
\newlength\textwidthtemp%
\newlength\textheightstd%
\newlength\textwidthstd%
\newlength\textheightold%
\newlength\textwidthold%
\newlength\tempheight%
\newlength\tempwidth%
\SetProtrusion{encoding={*},family={bch},series={*},size={6,7}}
              {1={ ,750},2={ ,500},3={ ,500},4={ ,500},5={ ,500},
               6={ ,500},7={ ,600},8={ ,500},9={ ,500},0={ ,500}}
\SetExtraKerning[unit=space]
    {encoding={*}, family={bch}, series={*}, size={footnotesize,small,normalsize}}
    {\textendash={400,400}, 
     "28={ ,150}, 
     "29={150, }, 
     \textquotedblleft={ ,150}, 
     \textquotedblright={150, }} 
\SetTracking{encoding={*}, shape=sc}{40}
\makeatletter
\let\blx@rerun@biber\relax
\makeatother
\usepackage{listings}
\usepackage{newtxmath}
\usetikzlibrary{external}
\pgfplotsset{
  grid style = {
   line width = 0.1pt
  }
}
				\newcommand{\disablewr}[1]{#1}%
				\newcommand{\newcommanddisw}[3]{\newcommand{#1}[1]{\disablewr{\textcolor{#2}{#3}}}}%
\renewcommand{\disablewr}[1]{}%
\definecolor{todocol}{named}{red}
\newcommanddisw{\todo}{todocol}{ToDo: #1}%
\definecolor{migucol}{named}{purple}%
\newcommanddisw{\migucom}{migucol}{{@}comment: #1}%
\newcommanddisw{\miguhigh}{migucol}{#1}%

\usepackage{lipsum}

\newcommand\blfootnote[1]{%
  \begingroup
  \renewcommand\thefootnote{}\footnote{#1}%
  \addtocounter{footnote}{-1}%
  \endgroup
}

\usepackage{siunitx}
\usepackage{xspace}

\definecolor{chhucol}{named}{blue}%
\newcommanddisw{\chhucom}{chhucol}{{@}comment: #1}%
\newcommanddisw{\chhuhigh}{chhucol}{#1}%

\definecolor{perocol}{named}{OliveGreen}%
\newcommanddisw{\perocom}{perocol}{{@}comment: #1}%
\newcommanddisw{\perohigh}{perocol}{#1}%

\definecolor{ruhacol}{named}{orange}%
\newcommanddisw{\ruhacom}{ruhacol}{{@}comment: #1}%
\newcommanddisw{\ruhahigh}{ruhacol}{#1}%

	\newcommand{\TempDisplayPreparation}{\disablewr{%
		\section{Draft-State: Comment Color Code}\noindent%
		\todo{Comments: ToDos}\nl%
		\migucom{Comments: Michael Gundall}\nl%

	}}%
\begin{document}%
%
\title{%
Integration of IEEE 802.1AS-based Time Synchronization in IEEE 802.11 as an Enabler for Novel Industrial Use Cases
}
%
\author{%
\IEEEauthorblockN{%
    Michael Gundall\IEEEauthorrefmark{1}, %
    Christopher Huber\IEEEauthorrefmark{1}, %
    and Sergiy Melnyk\IEEEauthorrefmark{1} %
    \\%
}%
\IEEEauthorblockA{%
    \IEEEauthorrefmark{1}German Research Center for Artificial Intelligence GmbH (DFKI), Kaiserslautern, Germany \\%
	\\%
    Email: %
        \{michael.gundall, christopher.huber, sergiy.melnyk
        \}@dfki.de
        \\%
}%
}%


%

%
%
%
%
%
%
%
%
\maketitle
%
%
%
%
%
\begin{abstract}%
Industry 4.0 introduces new use cases, with more and more mobile devices appearing in the industrial landscape. These applications require both new technologies and smooth integration into existing “brownfield" deployments. Emerging mobile use cases can be divided into optional mobile and mandatory mobile, where the first point considers the use of wireless communications due to soft criteria such as cost savings and the second means use cases that cannot be covered by wireline technologies due to their movement, such as AGVs. For most industrial applications, high determinism, E2E latency and synchronicity are most important. Therefore, we provide a common table, based on these requirements, listing both existing and emerging mobile use cases. Since time synchronization is particularly demanding for wireless use cases, we propose a concept for a simple but precise synchronization in IEEE 802.11 \gls{wlan} and a suitable integration using TSN in combination with OPC UA technology as examples. Furthermore, the concept is evaluated with the help of a testbed utilizing state-of-the-art hardware. This means that this concept can be directly applied in existing industry solutions. It can be shown that the concept is already suitable for a wide range of the mandatory mobile applications.
\end{abstract}%
%
%
%
%
%
\IEEEpeerreviewmaketitle
%
%
%
%
%
%
%
%
\section{Introduction}%
\label{sec:Introduction}


Many novel use cases are emerging in the context of Industry 4.0 \cite{8502649}. These use cases enhance the traditional applications in order to ensure the required flexibility of a smart manufacturing. One of the major differences is the requirement of wireless communications in order to allow the increasing number of mobile use cases. Tab. \ref{tab:Target use cases and selected requirement} sums up important use cases into classes and lists selected requirements, whereby the real-time classes are named from 1-3, or from A-C, dependant on the specific reference. In the following we use latter.\blfootnote{This is a preprint of a work accepted but not yet published at the 1st  Workshop on Next Generation Networks and Applications (NGNA). Please cite as: M. Gundall, C. Huber, and S. Melnyk: “Integration of IEEE 802.1AS-based Time Synchronization in IEEE 802.11 as an Enabler for Novel Industrial Use Cases”. In: 2020 1st  Workshop on Next Generation Networks and Applications (NGNA), 2020.}

Typical use cases that can be found in the remote control and monitoring use case class are predictive maintenance and those that are part of the \gls{ar} domain. These applications, which belong to the lowest real-time class A, require time synchronization better than 1~s. The challenges for these use cases are usually the data rates that need to be supported due to the number of sensor nodes or video transmissions and the coverage of a large area, but this is not the subject of this paper.
For mobile use cases, which are the main objective of the work in this paper and belong to the second class, the realization is particularly challenging, as they require wireless communication links with higher performance. Very challenging are those use cases where several mobile devices have a collaborative task, as here the most accurate time and state synchronization is required, where better synchronicity leads to faster robot interaction and thus to higher productivity. Typically, these use cases require a synchronicity of~$<$1~ms. 
Since use case class III has the highest requirements for \gls{e2e} latency and synchronicity, and cannot be addressed by current wireless communications, the use cases that belong to this class are based on wireline systems, such as industrial Ethernet. Here concepts for the combination of \gls{tsn} and \gls{5g} \cite{8731776}, as well as \gls{tsn} and \gls{wlan} \cite{mildner2019time,adame2019time} were introduced. The main arguments for replacing cables with wireless communication in this class are high cost savings \cite{7782431,mildner2019time,7883994}. Since the goal of our investigations is the use of existing hardware to enable mobile use cases, which necessarily require both wireless communication links and precise time synchronization, we will not deal with use case class III in our work.

Therefore, the following contributions can be found in this paper:
\textbf{
\begin{itemize}
   \item Integration of IEEE 802.1AS in IEEE 802.11 in order to fulfill the synchronicity imposed by industrial mobile use cases.
  \item Performance evaluation of the proposed concept based on a testbed. 
\end{itemize}
}

Accordingly, the paper is structured as follows: Sec. \ref{sec:Background} gives an overview about core technologies, while Sec. \ref{sec:Related Work} gives insights into related work on this topic. Details on the integration of IEEE 802.1AS and IEEE 802.11 are given in Sec. \ref{sec:Concept}. This is followed by a performance evaluation of the proposed concept based on a testbed consisting mainly of \gls{cots} hardware (Sec. \ref{sec:Testbed}). Finally, the paper is concluded in Sec. \ref{sec:Conclusion}.

\begin{table}[htbp]
\caption{Novel use cases and selected requirements \cite{8502649,7782431,8731776}}
\begin{center}
\begin{tabular*}{\columnwidth}{p{0.26\columnwidth}ccc}
\cline{1-3} 
\hline \hline
Use case class & \multicolumn{2}{c}{Requirements} & Real-time \\
\cline{2-3}
 & E2E latency & Synchronicity & class \\
\hline
(I) &  &  &  \\
Remote control, monitoring & 10-100 ms & $\leq$ 1 s & 1 / A \\
\hline
(II) &  &  &  \\
Mobile robotics, process control & 1-10 ms & $\leq$ 1 ms & 2 / B \\
\hline
(III) &  &  &  \\
Closed loop motion control & $<$ 1 ms & $\leq$ 1 $\mu$s & 3 / C \\
\hline \hline
\end{tabular*}
\label{tab:Target use cases and selected requirement}
\end{center}
\end{table}

\section{Core Technologies}%
\label{sec:Background}
In order to meet the requirements that are imposed by emerging mobile use cases, novel technologies are developed, whereby this section gives an overview of the most important technologies for this paper.


\subsection{IEEE Time-Sensitive Networking (TSN)}
\gls{tsn} being developed by the Time-Sensitive Networking task group of the IEEE 802.1 working group \cite{IEEETSN} define mechanisms for the time-sensitive transmission of data over deterministic Ethernet networks with respect to guaranteed \gls{e2e} latencies, reliability and fault tolerance \cite{8412459}. The use of IEEE 802 Ethernet in industrial applications that meet the requirements of industrial environments can also replace proprietary \gls{ie} protocols. Fort time synchronization the only the IEEE~802.1AS standard is relevant.


\subsection{OPC Unified Architecture (OPC UA)}
\gls{opcua} is the platform independent successor of the OPC standard, developed by the OPC Foundation \cite{IEC625411}. 
Major achievements of \gls{opcua} is the secure, easy and platform independent exchange of information between industrial devices. Furthermore, \gls{opcua} describes several communication protocols, e.g. for server-client communication, and \gls{pubsub} pattern. 

With this paper, \gls{opcua} \gls{pubsub} is very interesting because this pattern, described in part 14 of the specifications \cite{IEC625414}, allows many subscribers to register for a certain content. The message distribution includes both broker-based protocols, especially \gls{mqtt} and \gls{amqp}, and UADP, a custom \acrshort{udp}-based distribution based on the \acrshort{ip} standard for multicasting. Because of the advantages of sending real-time messages at the field level directly on the data link layer, Part 14 defines the transport of \gls{pubsub} messages based on Ethernet frames.

\subsection{IEEE 802.11}
\newcommand{\wlan}[1][]{IEEE\,802.11#1\xspace}
\wlan has basically two operating modes, infrastructure and ad-hoc mode. In ad-hoc mode, mobile devices, the so-called stations, transmit directly peer-to-peer. In infrastructure mode, stations communicate via an \gls{ap}. Since the infrastructure mode is used in almost all cases, both in the store and on the office floor, and our concept is only applicable to this mode, we will only discuss this mode in the following.

Moreover, \wlan defines a family of standards for \gls{wlan}, which target high data rate communication with wide coverage area for high number of stations. It includes a set of communication protocols for licence exempt bands of \SIlist{2.4; 5; 60}{\GHz} such as \wlan[a/b/g/n/ac/ad]. 
The current standard, also referred as Wi-Fi~5, utilises such techniques as \gls{ofdm} and \gls{mimo} in order to increase throughput. On \SI{2.4}{\GHz} frequency band, theoretical data rate of up to \SI{600}{Mbps} may be achieved with $4\times4$-\gls{mimo}. On \SI{5}{\GHz} frequency band, data rate of up to \SI{6.933}{Gbps} might be possible by means of $8\times8$-\gls{mimo} \cite{melnyk:2017}. Since Wave~2 extension,  the standard introduces \gls{mumimo} technique among others, which allows the \gls{ap} to transmit to several stations simultaneously. 
Furthermore, a new version of the standard, the \wlan[ax] or Wi-Fi~6, is recently emerged. Compared to the previous standard, its physical layer techniques further improve the data throughput. On the one hand, higher modulation schemes are available, and parallel uplink transmission is made possible by means of \gls{ofdma}. \gls{wlan} stations, which belong to the same network, carry out their communication  via an \gls{ap} by using a common channel. Based on a \gls{csma} scheme, any station  needs to ensure, that it would not cause interference with ongoing communication, prior to starting a transmission. The particular methods to realise \gls{csma} are defined by amendment \wlan[e], which enhances the initial access schemes with \gls{qos} capabilities by means of \gls{hcf} \cite{melnyk:2017}. 

On the one hand, \gls{hcf} defines a probabilistic \gls{edca} scheme. Prior to a transmission, a station has to generate a random back-off time period, which is a count down to start the transmission. In the case, some other transmission is detected during this period, the count down is paused until the channel is freed. This procedure lowers the collision probability as well as ensures fair channel access for any station. Furthermore, \gls{edca} introduces four \gls{qos} traffic classes, which are prioritised to each other. The higher the priority, the shorter back-off time is assigned to the traffic. In this way, the channel access probability is increased for higher priority traffic. In the case of the idle channel, the latency of below 10 ms could be achieved with \gls{edca} \cite{xiao2002throughput}. However, the raise of data traffic leads to significant raise of latency as well as the drop of throughput due to increasing number of collisions. Furthermore, it is not possible to provide any latency guarantee despite the traffic prioritisation mechanisms. On the other hand, \gls{hcca} provides a deterministic access scheme for \gls{wlan}. It introduces contention-free period, which is periodically advertised by the \gls{ap}. During this phase, \gls{ap} takes the control on channel access. By polling the stations accordingly, a scheduling approach can be realised.  Eventhough \gls{hcca} allows very flexible traffic scheduling, only a simple static scheduler is proposed by the standard. However, in \cite{melnyk2018hybrid} was shown, that by means of sophisticated scheduling schemes, the latency requirement of \SI{8}{\ms} could be guaranteed within a heterogeneous industrial environment.
Furthermore, the authors in \cite{ruscelli:2014} give a comprehensive overview on currently available \gls{hcca} scheduling algorithms and their performance.It should be mentioned, that the \gls{hcca} scheme is barely implemented. To the best of authors' knowledge, the only commercial solution available is iWLAN by Siemens [13]. It provides a proprietary iPCF functionality, which is capable of traffic scheduling for industrial applications.  Unfortunately, no reliable numbers on the performance could be found in the literature.

Last but not least, a \gls{tdma} approach in \gls{wlan} for automation purposes is described by \cite{wei:2013}. Instead of \gls{hcca} contention-free period, authors propose to introduce \gls{tdma} phase in order to provide dynamic scheduling of real-time traffic. The major advantages compared to \gls{hcca} are reduced protocol overhead as well as improved deterministic behaviour.

\section{Related Work}%
\label{sec:Related Work}
Precise clock synchronization in wireline systems is based on a constant time for data transmission between devices. Since wireless devices are often mobile and the propagation path of the communication signal changes during operation, this feature cannot be assumed for these systems. Therefore, the time synchronization of wireless systems requires other approaches. The different possibilities for time synchronization of wireless communications are described in \cite{9145977}, while \cite{7782431} focuses on \wlan \gls{wlan}. Furthermore, \cite{9145977} proposed a concept to adopt the state-of-the-art by using the received power of the station in order to estimate the distance to the \gls{ap}. In addition, \cite{7018946} focuses on the integration of \gls{ptp}, that has been defined in IEEE 1588 \cite{4579760} and \wlan using \gls{rbis} protocol. This method uses the broadcast character of wireless medium and is well suited for a simple but precise time synchronization of \wlan stations. Therefore, this method has also been adopted for the synchronization of \gls{3gpp} 4G and \gls{5g} systems \cite{gundall2020integration}. Moreover, \cite{gundall2020integration} used a IEEE~802.1AS Grandmaster as clock source to be \gls{tsn} conform and transmitted the time offset with \gls{opcua} \gls{pubsub} which is going to be the de-facto standard for application layer protocols in industrial automation.

\section{Integration of IEEE 802.11 with IEEE 802.1AS}%
\label{sec:Concept}
As already mentioned, most \gls{wlan} installations use the infrastructure mode, in which the stations do not communicate with each other directly, but via an \gls{ap}. To use the \gls{rbis} protocol for time synchronization of the stations, the following conditions must be satisfied:  
\begin{itemize}
    \item The message should arrive at each station at the same time
    \item Each message should have a unique identifier that is not repeatable
    \item The frequency of message transmission should be high
\end{itemize}

In addition to the user data that applications send from one station to another via an \gls{ap}, there are also control and management messages that are transmitted by the \gls{ap}, for example to share metadata. One of the management messages sent by each \gls{ap} is shown in the following listing:

\begin{lstlisting}
IEEE 802.11 Beacon frame
    Frame Control (2 Byte)
    Duration (2 Byte)
    Destination Address (6 Byte)
    Source Address (6 Byte)
    BSS ID (6 Byte)
    Seq-Ctl (2 Byte)
    Frame Body
        Timestamp (8 Byte)
        Beacon Interval (2 Byte)
        Capaility Info (2 Byte)
        ...
\end{lstlisting}
These messages are called beacon frames and contain the SSID of the \gls{ap}, the time interval of the transmission, and the timestamp of the beacon, i.e. the time that elapsed since the \gls{ap} was powered. By default the \gls{bi} is 100, i.e. 102.4 ms. If the synchronization should be improved, this value can easily be adopted, but with a higher number of management frames transmitted, the maximum data rate will be reduced. In a realistic industry landscape, not only one but multiple \glspl{ap} are in range of each station in order to guarantee seamless coverage. Thus a station usually receives beacon frames from multiple \glspl{ap}. To separate them, it is useful to filter by the BSS ID, which corresponds to the Mac address of the \gls{ap} and is also transmitted in the beacon frame. Because of the characteristics mentioned above, beacon frames are thus well suited for the \gls{rbis} protocol.

\begin{figure}[htbp]
\centerline{\includegraphics[width=\columnwidth]{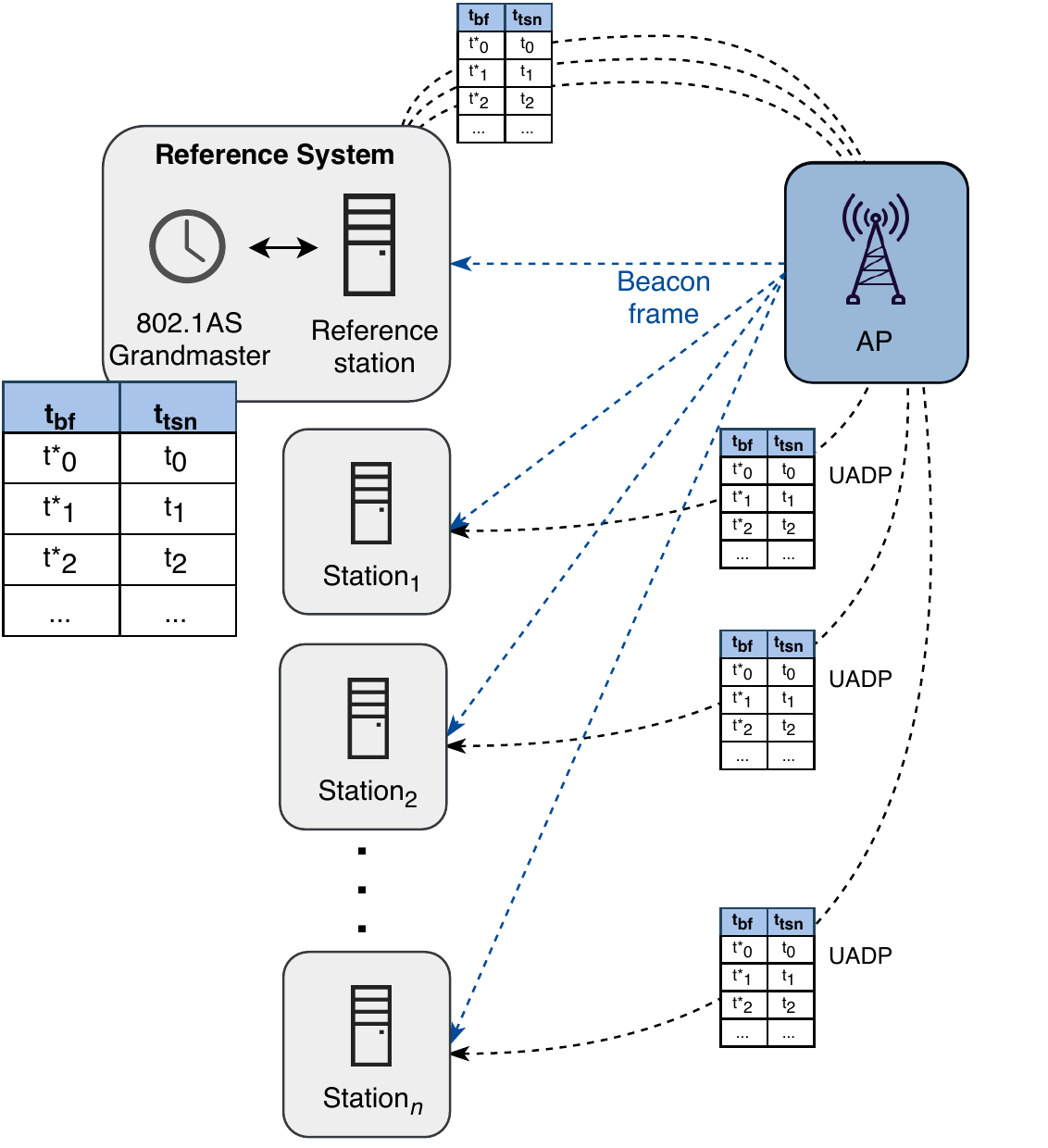}}
\caption{Concept for the distribution of the \gls{tsn} time in the \gls{wlan} network.}
\label{fig:concept}
\end{figure}

For the integration of time synchronization based on IEEE~802.1AS in \wlan, the concept shown in Fig. \ref{fig:concept} will be applied. It consists of an \gls{ap}, several stations, with one of them being called "Reference Station" and being part of the Reference System. What is special about this station is that it is connected to the wired \gls{tsn} network and so it cannot be mobile. In addition, this station is synchronized with \gls{tsn} time and must support IEEE 802.1AS. This synchronization is performed by the \gls{gm}, which can be any \gls{tsn} device. 

\begin{figure}[htbp]
\centerline{\includegraphics[scale=.83]{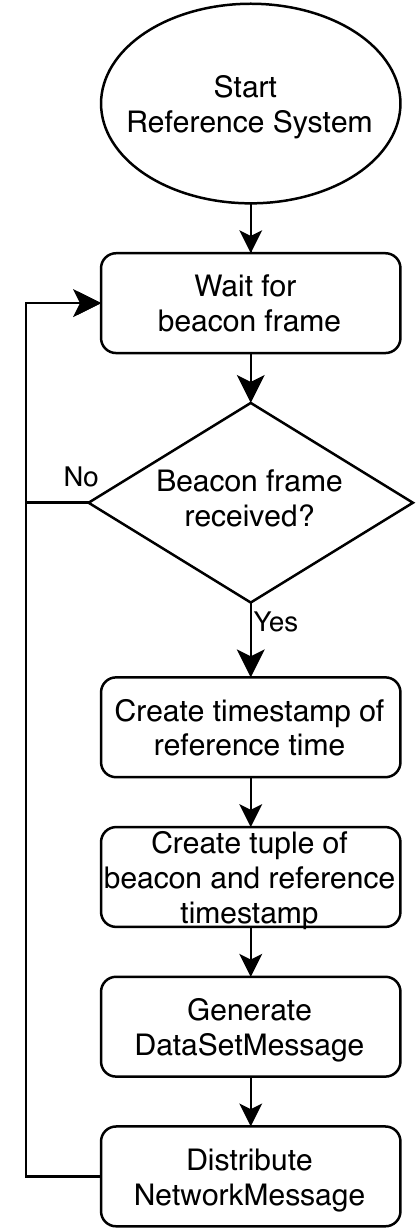}} 
\caption{Flow diagram of the Reference System}
\label{fig:Ref}
\end{figure}

\begin{figure}[htbp]
\centerline{\includegraphics[width=\columnwidth, trim = 130 380 165 330, clip]{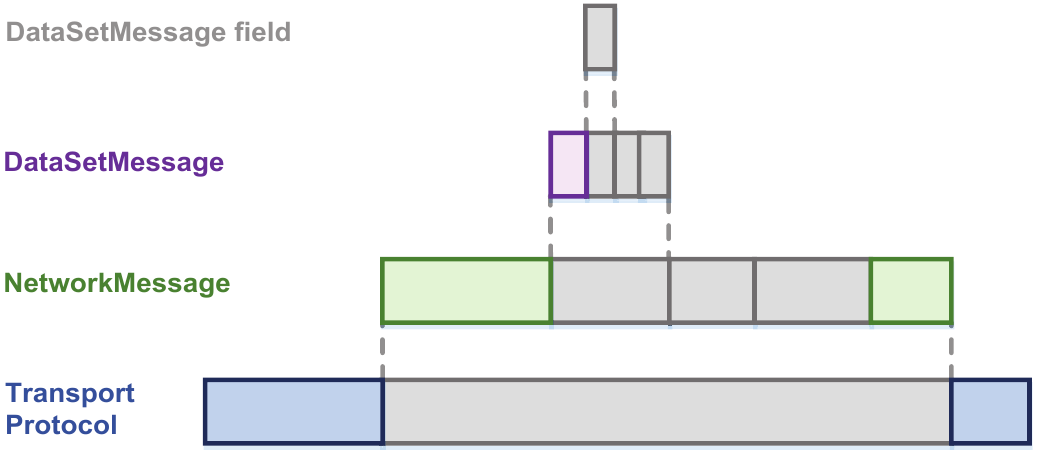}}
\caption{OPC UA PubSub message layers \cite{IEC625414}}
\label{fig:OPC UA PubSub message layers}
\end{figure}

In order to identify the correct offset to the \gls{tsn} time the Reference System pairs each incoming beacon frame timestamp with the corresponding \gls{tsn} timestamp, as shown in Fig. \ref{fig:Ref}.
Furthermore, the Reference Station sends this information to each station that has a subscription to this service. By using \gls{opcua} \gls{pubsub} for the distribution, it is possible to synchronize as many stations as are in range to the \gls{ap}, with the message layers shown in Fig. \ref{fig:OPC UA PubSub message layers}. 
The transport protocol used is UDP in combination with multicast, which means that each of the stations that have joined the multicast group receives the subscribed messages. If necessary, the transport protocol can also be changed from broker-less to broker-based, e.g. \gls{mqtt} or \gls{amqp}. It is also possible to distribute the messages via multicast based on layer 2. In addition there is the so-called \gls{opcua} \textit{NetworkMessage} which forms the payload of the UDP datagram, each \textit{NetworkMessage} having the \gls{opcua} specific header and footer and containing one or more \textit{DataSetMessages}, which in turn have so-called \textit{DataSetMessage} fields. In our case, the \textit{NetworkMessage} contains only one \textit{DataSetMessage}, which consists of its header and the following two \textit{DataSetMessage} fields: $t_{bf}$,  $t_{TSN}[t_{bf}]$. 

\begin{figure}[htbp]
\centerline{\includegraphics[scale=.83]{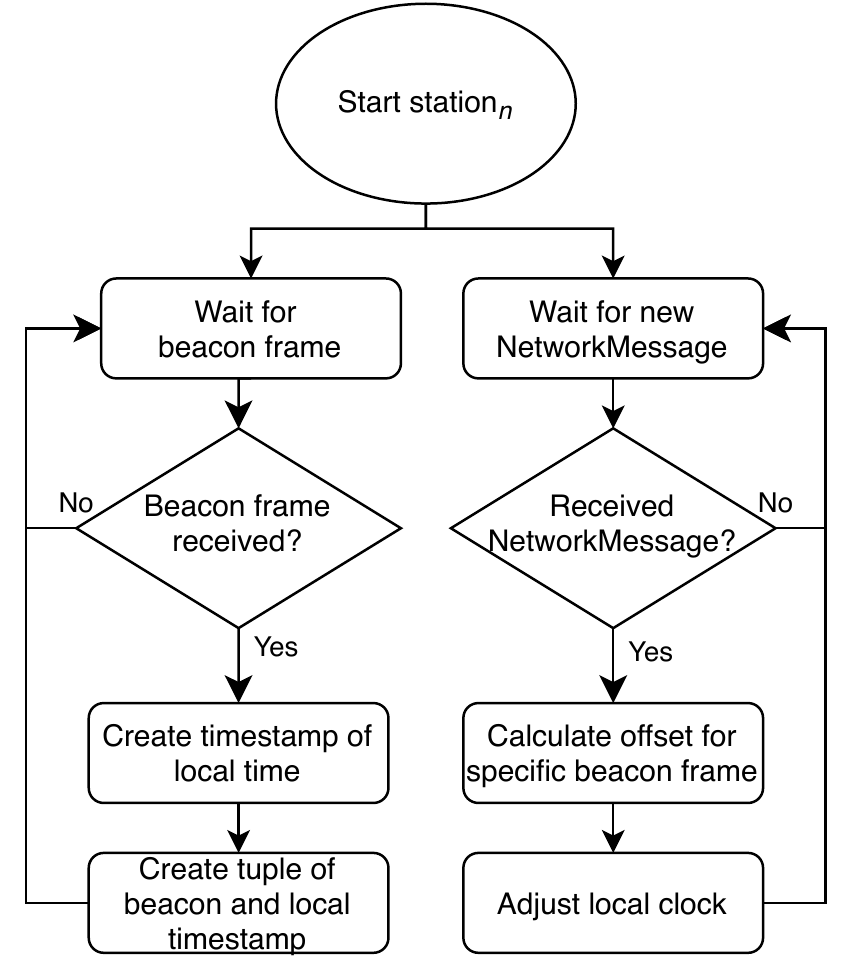}}
\caption{Flow diagram of a station that gets synchronized by the Reference System}
\label{fig:UE}
\end{figure}

The second workflow, which is similar for all other stations, is shown in Fig. \ref{fig:UE}. Here each station derives the tuples of its local time for each incoming beacon frame and the beacon frame timestamp. 
These tuples are used alongside the received tuples to calculate the offset and adjust the local clock accordingly. The formula for adjusting the local clock of the mobile stations is as follows, where 
$t_{TSN}[t_{bf}]$
is the time of the Reference System for a specific beacon frame, 
$t_{station}[t_{bf}]$
is the local time of the station that gets synchronized for the given beacon frame, and 
$t_{station}[current]$ 
 is the current time of the station: 
\begin{equation}
\label{eq:sn}
t_{TSN} = t_{TSN}[t_{bf}] - t_{station}[t_{bf}] + t_{ station }[current] 
\end{equation}

\section{Testbed \& Evaluation}%
\label{sec:Testbed}

\begin{figure}[htbp]
\centerline{\includegraphics[width=\columnwidth]{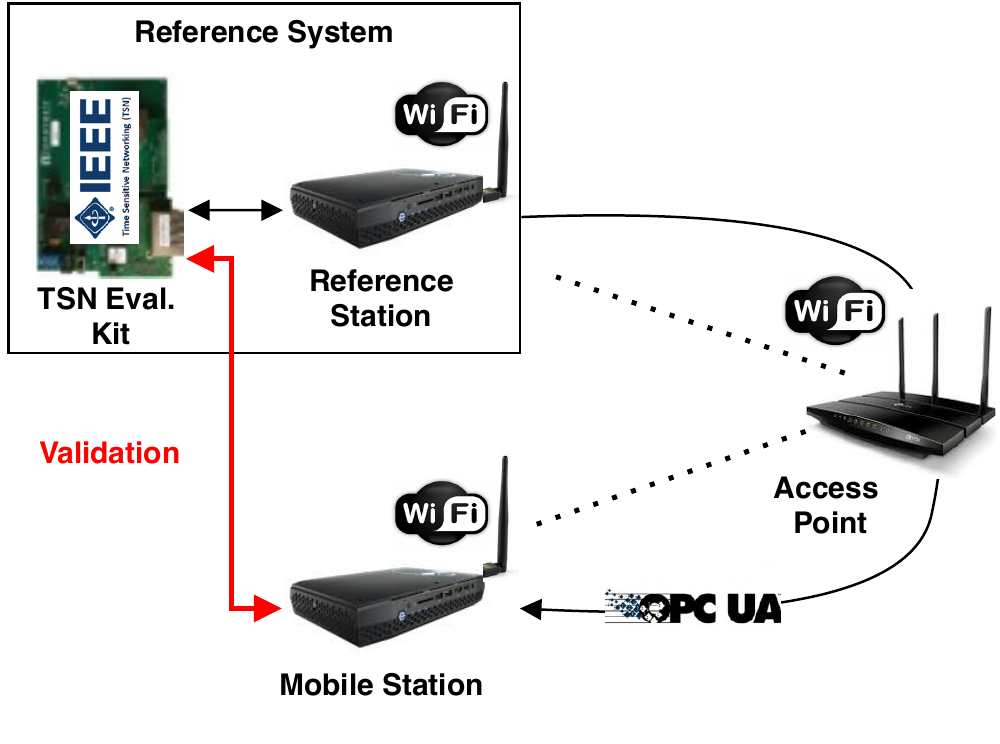}}
\caption{Testbed}
\label{fig:Testbed}
\end{figure}

This section aims to evaluate the proposed concept. Therefore, Fig. \ref{fig:Testbed} shows the testbed, on basis of which the evaluation has been done. It mainly consists of the \gls{cots} components, listed in Tab. \ref{tab:hardware} .

\begin{table}[htbp]
\caption{Hardware configurations}
\begin{center}
\begin{tabular*}{\columnwidth}{|p{0.26\columnwidth}|c|p{0.5\columnwidth}|}
\cline{1-3} 
\textbf{\textit{Equipment}} & \textbf{\textit{QTY}} & \textbf{\textit{Specification}}\\
\cline{1-3} 
Mini PC & 2 & Intel Core i7-8809G, 2x16 GB DDR4, Intel i210-AT \& i219-LM Gibgabit NICs, Ubuntu 18.04.3 LTS 64-bit, \linebreak Linux 4.18.0-18-lowlatency  \\
\cline{1-3} 
Wi-Fi Adapter & 2 & USB, IEEE~802.11a/g/b/n/ac, Wi-Fi 5\\
\cline{1-3} 
Wi-Fi Router & 1 & IEEE~802.11a/g/b/n/ac, Wi-Fi 5\\
\cline{1-3}
TSN Evaluation Kit & 1 & RAPID-TSNEK-V0001, IEEE~802.1AS \\
\cline{1-3} 
\end{tabular*}
\label{tab:hardware}
\end{center}
\end{table}

It consists of two identical mini \glspl{pc}, that are connected wireless to a Wi-Fi router that serves as \gls{ap}. In order to receive each beacon frame, the \gls{wlan} network interfaces have to be set in "monitor mode" by using aricrack-ng module 
\cite{aircrack}. Afterwards, the channel can be monitored, but the IPv4 connectivity gets lost. In order to transfer the \textit{NetworkMessages}, an additional Wi-Fi adapter was added per mini \gls{pc} via USB. Next, the mini \gls{pc} that serves as Reference Station is connected to \gls{tsn} Evaluation Kit. It supports the IEEE 802.1AS and 802.1AS-REV specifications and can consequently serve as \gls{gm} for other \gls{tsn} devices. Furthermore, Linux PTP 
is a free and open source software \gls{ptp} implementation that complies with the IEEE 1588 standard \cite{cochran2015linux}. This implementation is one of the most frequently used. Besides aiming to provide a robust implementation of the standard Linux \gls{ptp} tries to make use of the most relevant and modern \glspl{api} offered by the Linux kernel. The Linux \gls{ptp} project provides several executables to run two-stage synchronization mechanism. The one which was used in our testbed is \textit{ptp4l}. 

The \textit{ptp4l} tool synchronizes the \gls{ptp} hardware clock with the master clock in the network. If there is no \gls{ptp} hardware clock, it automatically synchronizes the system clock with a master clock using software timestamps. As extension, the tool supports the IEEE 802.1AS specification for \gls{tsn} end stations, by using the \gls{gptp} configuration file, which modifies the default procedure of the executable.

\begin{figure}[htbp]
\resizebox{\columnwidth}{!}{%
\input{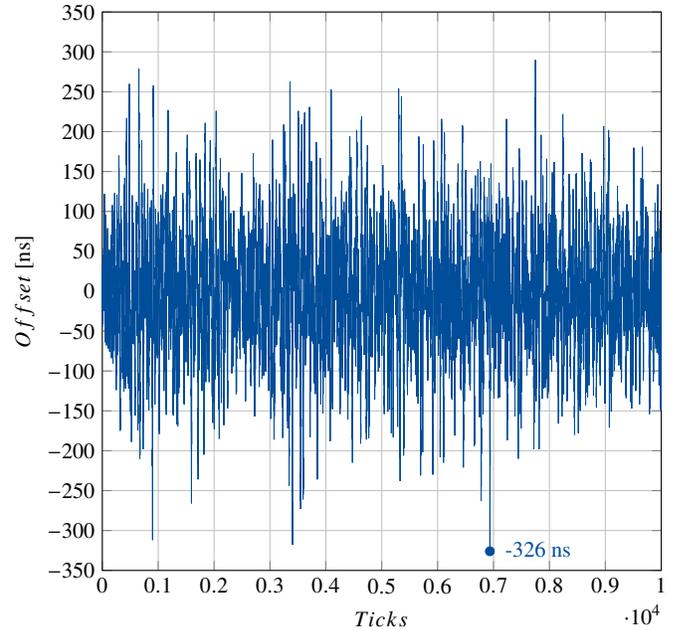}}
\caption{Synchronization accuracy between \gls{tsn} Evaluation Kit and Intel NUC mini PC for a sync interval of 31.25ms (2\textsuperscript{-5}s).}
\label{fig:gPTP}
\end{figure}

As shown in Fig. \ref{fig:gPTP} a synchronicity of $\pm$ 350ns between \gls{tsn} Evaluation Kit and the Reference Station can  be reached, by using the minimum sync interval of 31.25ms (2\textsuperscript{-5}s).

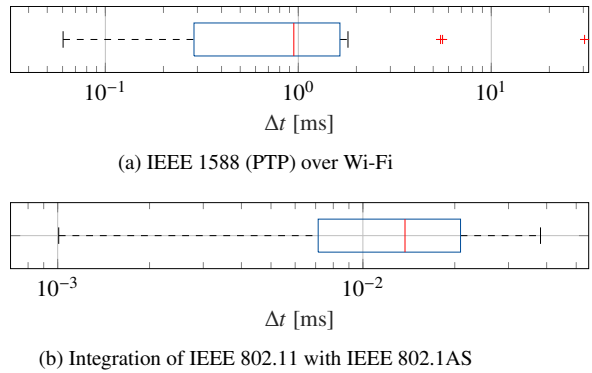
\begin{figure}[htbp]
	\centering
		\subfloat[IEEE 1588 (PTP) over Wi-Fi]{\resizebox{\columnwidth}{!}{%
%
%
\definecolor{mycolor1}{rgb}{0.00000,0.3,0.6}%
\begin{tikzpicture}

\begin{axis}[%
width=3.5in,
height=0.4in,
scale only axis,
unbounded coords=jump,
xmode=log,
xmin=-1.45863275,
xmax=31.95995575,
xlabel style={font=\color{white!15!black}},
xlabel={$\Delta t$ [ms]},
ymin=1.7,
ymax=2.3,
ytick={1},
yticklabels={\empty},
ylabel style={font=\color{white!15!black}},
ylabel={~},
axis background/.style={fill=white},
xmajorgrids,
ymajorgrids,
]
\addplot [color=black, dashed, forget plot]
  table[row sep=crcr]{%
1.6432615	2\\
1.808659	2\\
};
\addplot [color=black, dashed, forget plot]
  table[row sep=crcr]{%
0.060394	2\\
0.2879375	2\\
};
\addplot [color=black, forget plot]
  table[row sep=crcr]{%
1.808659	1.925\\
1.808659	2.075\\
};
\addplot [color=black, forget plot]
  table[row sep=crcr]{%
0.060394	1.925\\
0.060394	2.075\\
};
\addplot [color=mycolor1, forget plot]
  table[row sep=crcr]{%
0.2879375	1.85\\
1.6432615	1.85\\
1.6432615	2.15\\
0.2879375	2.15\\
0.2879375	1.85\\
};
\addplot [color=red, forget plot]
  table[row sep=crcr]{%
0.9475505	1.85\\
0.9475505	2.15\\
};
\addplot [color=black, draw=none, mark=+, mark options={solid, red}, forget plot]
  table[row sep=crcr]{%
5.459047	2\\
5.591106	2\\
30.440929	2\\
};
\end{axis}
\end{tikzpicture}%
}\label{fig:Sub_PTP}}

		\subfloat[Integration of IEEE 802.11 with IEEE 802.1AS]{\resizebox{\columnwidth}{!}{%
%
%
\definecolor{mycolor1}{rgb}{0.00000,0.3,0.6}%
\begin{tikzpicture}

\begin{axis}[%
width=3.5in,
height=0.4in,
scale only axis,
unbounded coords=jump,
xmode=log,
xlabel style={font=\color{white!15!black}},
xlabel={$\Delta t$ [ms]},
ymin=0.7,
ymax=1.3,
ytick={1},
yticklabels={\empty},
ylabel style={font=\color{white!15!black}},
ylabel={~},
axis background/.style={fill=white},
xmajorgrids,
ymajorgrids,
]
\addplot [color=black, dashed, forget plot]
  table[row sep=crcr]{%
0.0208905   1\\
0.038196    1\\
};
\addplot [color=black, dashed, forget plot]
  table[row sep=crcr]{%
0.001008	1\\
0.00713775  1\\
};
\addplot [color=black, forget plot]
  table[row sep=crcr]{%
0.038196 	0.925\\
0.038196 	1.075\\
};
\addplot [color=black, forget plot]
  table[row sep=crcr]{%
0.001008	0.925\\
0.001008	1.075\\
};
\addplot [color=mycolor1, forget plot]
  table[row sep=crcr]{%
0.00713775	0.85\\
0.0208905	0.85\\
0.0208905	1.15\\
0.00713775	1.15\\
0.00713775	0.85\\
};
\addplot [color=red, forget plot]
  table[row sep=crcr]{%
0.0137405	0.85\\
0.0137405	1.15\\
};
\end{axis}
\end{tikzpicture}%
}\label{fig:Sub_SFN}}

\caption{Readings of the measurements for the evaluation of our concept compared to the state of the art solution.}
\label{fig:results}
\end{figure}

In order to evaluate the quality of the time synchronization, the time difference between both stations has to be identified. For this reason, the \gls{tsn} Evaluation Kit is also connected to the second station and synchronizes the hardware clock of one of the built in network interfaces. Now, the time difference between the internal clock and the hardware clock can be measured. In addition to our implementation, the state-of-the art for time synchronization of wireline systems has also been applied to the testbed. The results of both measurements are shown in Fig.~\ref{fig:results}. 

In the first figure (Fig. \ref{fig:Sub_PTP}), in which the IEEE 1588 protocol is transmitted directly via Wi-Fi, it can be seen that there are at least some use cases of class II can be fulfilled by the synchronization. This is also reflected in the median value, which is with $\approx$0.95~ms below the limit of~$<$1~ms. However, there are many values in the 1-2~ms range derived from the already mentioned drawbacks of this method and even some values that exceed this value. The maximum, which is above 30~ms, is not acceptable for use case class II. 

Fig. \ref{fig:Sub_SFN}, which is based on the method presented in this paper, has a median of 13~µs and a maximum value of 38~µs. Thus, this approach is suitable to fulfill the different use cases of group II and III. This means that the required time synchronization of all mandatory mobile use cases required for Industry 4.0 can be achieved by this approach using \gls{cots} hardware that can be found in existing industrial facilities.

\section{Conclusion}%
\label{sec:Conclusion}
In this work, we proposed a concept for the time synchronization of \wlan-based \gls{wlan} and the corresponding integration into IEEE~802.1AS systems. Using a testbed consisting of COTS Wi-Fi components, it can be shown that this solution can provide better time synchronization than if wired algorithms were applied to the wireless communication. In addition, the results show that a wide range of so-called mandatory mobile use cases, such as AGVs that transport a workpiece together, can be covered by this approach.




\printbibliography%

\TempDisplayPreparation
\end{document}